\documentclass[manuscript,screen]{acmart}

\usepackage{caption}
\usepackage{subcaption}
\usepackage{tabularx}
\usepackage{colortbl}
\usepackage{xcolor}
\usepackage{longtable}
\usepackage{booktabs}

\usepackage{float}
\floatstyle{plaintop}
\restylefloat{table}

\AtBeginDocument{%
  \providecommand\BibTeX{{%
    \normalfont B\kern-0.5em{\scshape i\kern-0.25em b}\kern-0.8em\TeX}}}

\setcopyright{acmcopyright}
\copyrightyear{2024}
\acmYear{2024}
\acmDOI{XXXXXXX.XXXXXXX}

\begin{document}

\title[IDC x CA Design Challenges]{Exploring Links between Conversational Agent Design Challenges and Interdisciplinary Collaboration}


\author{Malak Sadek}
\email{m.sadek21@imperial.ac.uk}
\orcid{0000-0001-8284-890X}
\author{Céline Mougenot}
\email{c.mougenot@imperial.ac.uk}
\affiliation{%
  \institution{Imperial College London}
  \streetaddress{South Kensington}
  \city{London}
  \country{United Kingdom}
  \postcode{SW7 2BX}
}

\renewcommand{\shortauthors}{Trovato and Tobin, et al.}

\begin{abstract}
Recent years have seen a steady rise in the popularity and use of Conversational Agents (CA) for different applications, well before the more immediate impact of large language models. This rise has been accompanied by an extensive exploration and documentation of the challenges of designing and creating conversational agents. Focusing on a recent scoping review of the socio-technical challenges of CA creation, this opinion paper calls for an examination of the extent with which interdisciplinary collaboration (IDC) challenges might contribute towards socio-technical CA design challenges. The paper proposes a taxonomy of CA design challenges using IDC as a lens, and proposes practical strategies to overcome them which compliment existing design principles. The paper invites future work to empirically verify suggested conceptual links and apply the proposed strategies within the space of CA design to evaluate their effectiveness.

\end{abstract}





\maketitle

\section{Introduction}
Offering a more human-like and interactive experience, the rise in popularity for Conversational Agents (CA) in recent years can be seen in research projects, industry-based applications, and commercial products. This increased use comes with increased research into how this technology is designed and built, and the challenges that hinder those practices. Specifically, \cite{heuer2022} conducted a scoping review of the socio-technical challenges and issues relating to CA creation both cited in research and mentioned by practitioners during semi-structured interviews. 

\bigskip
\noindent The aim of this opinion paper is to suggest an exploration of the links between CA design challenges presented by \cite{heuer2022} in their scoping review and interdisciplinary collaboration (IDC) challenges. We begin to explore this link by taxonimizing the CA design challenges using IDC as a lens and proposing practical strategies to compliment existing design principles in order to address them. CA design challenges were taxonomized into two groups: team culture challenges and team member challenges. These two groups drew upon several conceptual similarities to groupings of IDC challenges that are either based on communication, or the coming together of different disciplines. Following, two strategies were proposed, drawing upon work from the more mature fields of IDC and CSCW and applied to the domain of CA design. The proposed strategies (i) target team member challenges by focusing on facilitating communication through the use of boundary objects to unify language and knowledge across the team, and (ii) target team culture challenges by establishing team harmony through appropriate mindsets and practices, as well as through a focus on stakeholder values. The study concludes by outlining the future empirical work needed to validate the effectiveness of these strategies in having a positive influence on the challenges presented. 

\bigskip
\noindent While \cite{heuer2022} present a series of resulting design principles derived from the issues they collate, design principles, similar to recommendations and guidelines, come with a variety of criticisms and limitations \cite{gregor2020,iivari2021}. To overcome certain limitations, such as the abstract or generic nature of recommendations and their ambiguity and difficulty in on-ground implementation \cite{elshan2022}, the strategies presented based on this study's taxonomy aim to compliment \cite{heuer2022}'s design principles with more practical approaches that may prove beneficial in utilizing established techniques to overcome some of the CA design challenges reviewed. We invite future work to verify the links suggested by the proposed taxonomy and empirically validate proposed strategies in order to assess their effectiveness in alleviating or mitigating some of the challenges mentioned. 

\bigskip
\noindent This consultation of literature on interdisciplinary collaboration (IDC) is not to suggest that all challenges are solely caused by interdisciplinary collaboration. This hypothesis would be unjustified for several reasons; the two most prominent being that (i) not all conversational agents are developed in interdisciplinary settings even if many are, and (ii) empirical work would be needed to verify a link of causality between the two. Instead, given the strong recommendations to develop these technologies in interdisciplinary settings, the consultation is an investigation into how to address existing socio-technical challenges for CA design when taking into account their development within interdisciplinary settings. It serves as an aid when taxonomizing these challenges, and the strategies derived act as proposals for adapting the more abstract field of interdisciplinary collaboration to the practical and concrete steps, stakeholders, and activities of CA design. 

\bigskip
\noindent Accordingly, after providing needed background information, the paper begins by introducing the socio-technical challenges of CA design presented by \cite{heuer2022}. Following, a taxonomy of these challenges is suggested and the corresponding proposed strategies are laid out. The roots of these strategies lies in the fields of IDC and Computer Supported Collaborative Work (CSCW), which has a long history of addressing and supporting IDC in technology-based settings. Finally, future work needed to validate these proposed strategies empirically is outlined.

\section{Background}

\subsection{Conversational Agents}

In this study, a conversational agent is taken to be any system which provides a ``conversational experience” \cite{delloiteCA} to its users. Such a basic definition abstracts several dimensions such as input/output modalities, interface types, application domain or target users. By abstracting these dimensions, all conversational agents, ranging from text-based FAQ chatbots to highly advanced embodied assistants, are included in the scope of investigation. This abstraction is key since most reviews of CA design challenges make no differentiation between different aspects of the agents or position different dimensions as contributing specifically to any of the challenges. 

\bigskip
\noindent CAs are currently being used in a variety of domains \cite{laumer2019,motger2021} including healthcare \cite{fadhil2018,piau2019}, education \cite{bayne2015,wambsganss2020}, culture \& tourism \cite{kasinathan2020,nica2018}, and more social use-cases such as coding partners \cite{kuttal2020} or co-creative partners \cite{rezwana2022}. Many CAs rely both on programming and artificial intelligence, meaning that several disciplines and considerations need to be involved regarding both aspects. In fact, conversational AI lies at the intersection of several concepts such as computer programming, artificial intelligence, conversational design, interaction design, and ethics \cite{rasaDemocratise}.

\bigskip
\noindent While AI-based CAs share several shortcomings and design challenges with other AI-based systems, such as biases arising from training data \cite{ainow2019,byrd2020,folstad2021}, many of the challenges of designing and creating CAs are unique to this technology. Due to their conversational nature, CAs can act as a form of social or service actors in several contexts, especially when factors such as personalities and autonomy are added \cite{lewandowski2022}. This form of presence, which falls somewhere in the middle of human and machine, and the advent of public, general-use CAs such as ChatGPT, opens up new use-cases, interactions, and relationships \cite{seeber2020}. It also requires interdisciplinarity and diversity to capture different perspectives and experiences, and much deeper considerations of human values and ethics, especially as these actors begin getting involved in relationship building \cite{sakai2012} and task collaboration \cite{kuttal2020} with users. 

\bigskip
\noindent These challenges and conditions necessitate that CAs be built in more interdisciplinary settings, perhaps more so than other types of AI-based systems, which already commonly takes place and is highly recommended by several experts \cite{naganathan2021,motger2021}. Accordingly, the taxonomy established and strategies proposed are in-part derived from literature on interdisciplinary collaboration in order to harness the power of existing research within this comparatively more established field and adapt its learnings to CA design, where the infancy of the field and limited specialized knowledge might be exacerbating already existing challenges of working in interdisciplinary settings. 

\bigskip
\noindent The focus here is on the interdisciplinary nature of CA design. Unlike many other AI-based systems, creating a CA is a collaborative process that involves technical stakeholders (e.g. developers, integration engineers, etc.), business stakeholders (e.g. product owners, marketing, legal, etc.), designers (e.g. UX designers, conversation designers, etc.), domain experts, end-users, and a variety of other disciplines such as psychologists, editors, user researchers, etc. \cite{pearl2018}. The CA design process is a cross-disciplinary collaborative endeavor that benefits from divergent thinking between business stakeholders who provide the `why' and `what', designers who provide conversation flows and dialogues, domain experts who provide content and context, developers who provide technical constraints, and end-users who can find holes in the resulting iterative CA prototypes and refine it before development even begins \cite{mixter2019}.

\bigskip
\noindent Recently, there have been calls for having even more interdisciplinary collaboration throughout the CA design process \cite{naganathan2021,motger2021}. There have also been calls for especially leveraging domain expert and end-user knowledge \cite{naganathan2021,cambre2020,diederich2022} to help them foster a sense of ownership and control over CAs \cite{walby2019} and mitigate various organizational \cite{goebel2017}, and user risks \cite{ratner2021a,ratner2021b}. Within industry-based settings, it is viewed as crucial that designers are aware of technical constraints upfront and can design for them, and that other stakeholders understand designs early on so they can plan for them \cite{pearl2018,mixter2019}. This is not only an industry-based reality however, research-based design processes for CAs also explicitly account for engaging with these various stakeholders and understanding their needs and inputs \cite{wolff2022}. 

\bigskip
\noindent Overall, when designing CAs, IDC is in many cases a reality, but also a recommended best practice, and thus it becomes imperative to understand how this style of working contributes towards challenges in designing and creating CAs and how strategies to manage it can be applied to the field.

\subsection{Interdisciplinary Collaboration}

Despite the global calls for interdisciplinarity across a number of domains catapulting it into `buzzword' status, interdisciplinary work has been taking place for a long time. While the term `interdisciplinary studies' first appears around the 20th century, evidence suggests that civilizations as old as the ancient Egyptians and Greeks were involved in interdisciplinary work and research \cite{Al-Saleem2018}. Any activity can be described as interdisciplinary so long as the tools, knowledge, or frameworks it uses span across two or more disciplines \cite{nap2005,porter2009}; where a discipline is an academic area that has its own established way of generating knowledge, asking questions and seeking answers to them (i.e. its own tools, frameworks, models, terms, and so on) \cite{stephenson2010,alkindi2018}. To better understand the challenges of reconciling different disciplines, it is useful to contextualise the fact that disciplines also have their own ways of viewing the world \cite{klein1990,stephenson2010,eigenbrode2007} and their own scientific communities which members belong to and identify with \cite{aldrich2014,weingart2010,repko2012}. 

\bigskip
\noindent Interdisciplinarity leverages the strengths and the arsenal of resources available to the set of disciplines involved, resulting in a sum that is much more powerful and holistic than any of the single disciplines could have produced on their own \cite{LauraMeagher2011,Khamdamova2016,antonic2021}. It draws upon the perspectives, knowledge, tools, and methodological approaches of several disciplines \cite{winowiecki2011}, and allows for highly enhanced forms of collaboration and problem-solving \cite{winowiecki2011,Khamdamova2016,malone2022}. Because of these massive benefits, there is a lot of research on interdisciplinary collaboration and communication in critical applications such as healthcare and aviation \cite{Kuziemsky2009}. These benefits can also be extended to the domain of AI-based systems, such as CAs, helping to understand and reflect upon the wide-reaching and critical implications of these systems on various stakeholders. 

\bigskip
\noindent By having experts representing different disciplines come together, reach a state of shared understanding and agreement, and collaborate with each other to reach a common goal; existing myths, preconceptions, and biases about various disciplines can be dispelled \cite{LauraMeagher2011}. These biases and preconceptions are are commonly cited as one of the biggest hurdles to effective interdisciplinary communication \cite{norman2002,Rosenberg2011,Hayes2020}.

\bigskip
\noindent While the benefits of interdisciplinary work are numerous, this type of work also comes with a variety of challenges which have been well-documented. Nevertheless, several strategies have been developed to address many of these challenges. By leveraging established knowledge from this more mature field, the relatively juvenile space of CA design can benefit greatly in terms of beginning to counter some of the challenges it faces. The next section thus explores common socio-techincal challenges of CA design and then presents corresponding or similar challenges in the space of IDC. Finally, the two sets of challenges are explicitly mapped before moving on to leverage corresponding strategies for IDC challenges which can also be applied to similar CA challenges.

\subsection{Conversational Agents Design Challenges}

\cite{heuer2022}'s sweeping survey of studies revolving around conversational agents, coupled with in-depth interviews with experts involved with CAs, has comprehensively highlighted several recurring socio-technical issues within the domain of conversational agent design. Many of those issues have also been mentioned in a number of other studies in the field.

\bigskip
\noindent Table \ref{tab:33-challenges} summarises the thirteen socio-technical challenges that arise within teams creating CAs as described by \cite{heuer2022}. 

\begin{table}[]
\centering
\resizebox{\textwidth}{!}{%
\begin{tabular}{|c|l|}
\hline
\textbf{Challenge} & \multicolumn{1}{c|}{\textbf{Description}} \\ \hline
\textbf{\begin{tabular}[c]{@{}c@{}}Long-term vision and \\ roadmap\end{tabular}} & \begin{tabular}[c]{@{}l@{}}Lack of long-term commited vision and roadmap resulting in a \\ lack of resources and support allocated.\end{tabular} \\ \hline
\textbf{\begin{tabular}[c]{@{}c@{}}Expectations of novel \\ information systems\end{tabular}} & \begin{tabular}[c]{@{}l@{}}Insufficient or inaccurate knowledge, expectations, readiness, \\ acceptance, and skills regarding CAs across an organisation.\end{tabular} \\ \hline
\textbf{\begin{tabular}[c]{@{}c@{}}Release-rush \\ atmosphere\end{tabular}} & \begin{tabular}[c]{@{}l@{}}Underestimation of the preparation effort needed to produce \\ a sufficiently mature CA leading to a premature deployment \\ and thus long-term non-use.\end{tabular} \\ \hline
\textbf{\begin{tabular}[c]{@{}c@{}}Disregard of \\ underlying influences\end{tabular}} & \begin{tabular}[c]{@{}l@{}}Underestimation of the legal, ethical, and organisational issues \\ involved when creating CAs.\end{tabular} \\ \hline
\textbf{\begin{tabular}[c]{@{}c@{}}Integration and \\ modernisation of IT \\ landscape\end{tabular}} & \begin{tabular}[c]{@{}l@{}}CAs are technically detached from structures and architectures\\ within existing IT infrastructures.\end{tabular} \\ \hline
\textbf{\begin{tabular}[c]{@{}c@{}}Integration into work \\ structures and \\ processes\end{tabular}} & CAs are detached from business workflows and processes. \\ \hline
\textbf{\begin{tabular}[c]{@{}c@{}}Lack of new \\ responsibilities \\ and freedoms\end{tabular}} & \begin{tabular}[c]{@{}l@{}}Lack of continuous involvement of diverse company stakeholders\\ and the creation of suitable roles needed for further development.\end{tabular} \\ \hline
\textbf{\begin{tabular}[c]{@{}c@{}}Underestimation of \\ required competences\end{tabular}} & \begin{tabular}[c]{@{}l@{}}Underestimation of the expertise and competence needed for CA\\ development causes potential lock-ins to vendors and frameworks.\end{tabular} \\ \hline
\textbf{\begin{tabular}[c]{@{}c@{}}Distributed \\ knowledge in expert \\ domains\end{tabular}} & \begin{tabular}[c]{@{}l@{}}Lack of involvement of domain experts to support CA use-cases\\ and no training or diffusion of knowledge to other stakeholders.\end{tabular} \\ \hline
\textbf{\begin{tabular}[c]{@{}c@{}}Data availability and \\ natural language \\ processing-conformity\end{tabular}} & \begin{tabular}[c]{@{}l@{}}Underestimation of data management practices needed to produce\\ high-quality datasets for training.\end{tabular} \\ \hline
\textbf{\begin{tabular}[c]{@{}c@{}}Continuous training \\ and maintenance\end{tabular}} & \begin{tabular}[c]{@{}l@{}}Knowledge, technology and data are not kept up to date to ensure\\ contiuous development and training.\end{tabular} \\ \hline
\textbf{\begin{tabular}[c]{@{}c@{}}Continuous \\ monitoring and \\ visualisation\end{tabular}} & \begin{tabular}[c]{@{}l@{}}CA deployment lacks continuous monitoring to highlight\\ organisational benefits leads to low acceptance and participation.\end{tabular} \\ \hline
\textbf{\begin{tabular}[c]{@{}c@{}}Continuous \\ improvement culture\end{tabular}} & \begin{tabular}[c]{@{}l@{}}Poor feedback and communication culture in the organisation\\ hinders the dissemination of diverse knowledge needed throughout\\ CA development.\end{tabular} \\ \hline
\end{tabular}%
}
\bigskip
\caption{A summary and description of the socio-technical challenges presented by and adapted from \cite{heuer2022}.}
\label{tab:33-challenges}
\end{table}

\section{Proposed Taxonomy of Conversational Agent Design Challenges}

The challenges reviewed exist on two levels. On one level, some challenges are caused by differences between team members, such as having no unified vision or goals, no knowledge distribution, or having differences in expectations. These can often be caused by early-stage issues in project management, such as a lack of clear communication or transparency in the earlier stages of a CA design project, and might be more straightforward to correct.

\bigskip
\noindent On another level, other challenges arise from issues in broader team or even disciplinary cultures, such as a mismatch of team roles and responsibilities, the lack of focus on maintenance and monitoring, and the overlooking of socio-technical factors. These challenges require a deeper change and a more intentional adoption of different strategies to help fortify against and mitigate these problems. The strategies needed must aim to shift team culture by encouraging practices that go beyond simply ensuring members are on the same page and instead build up collaborative practices and traditions for healthy team functioning. 

\bigskip
\noindent A similar distinction can be found in challenges interdisciplinary collaboration in general. Some challenges are due to more micro-level, communication-based problems between team members, such as difficulty reaching a shared understanding or shared goals \cite{norman2002,LauraMeagher2011,winowiecki2011,antonic2021}, difficulty reaching a shared language for communication \cite{winowiecki2011,Rosenberg2011,antonic2021}, and miscommunication and lack of clarity when sharing various ideas and artifacts \cite{Khamdamova2016,winowiecki2011}. Communication-based challenges tend to arise during interactions between practitioners where there is no shared vision, no clear communication or no common language or shared knowledge repository. It is worth noting that many challenges which occur at this level are challenges that arise with teamwork in general and are not only specific to interdisciplinary collaboration.

\bigskip
\noindent Conversely, other challenges occur more at a macro-level when different disciplines come together. These include methodological confusion or mismatch \cite{crease2010,stephenson2010,LauraMeagher2011,smith2017,Rosenberg2011,siraj2018,macleod2018}, differences in expectations across disciplines regarding level of rigor or evaluation methods \cite{LauraMeagher2011,eigenbrode2007,Rosenberg2011,Khamdamova2016,sleeswijk2018,siraj2018}, difficulty fostering collective responsibility \cite{defila2006,stephenson2010,LauraMeagher2011,smith2017,antonic2021}, and biases and preconceptions about other disciplines \cite{norman2002,stephenson2010,Rosenberg2011,repko2012,xinlan2022}. These challenges occur at a more macro-level because of differences in methodologies, rigor, expectations and so on across disciplines, as well as biases and preconceptions between them. The resulting divide between disciplines can be so severe that it has caused researchers to picture different disciplines as being like different cultures \cite{abdullahi2018,snow2020,smith2017,crease2010} (in line with theoretical concepts such as ``epistemic cultures" \cite{knorr1999}) or different ``thought worlds" \cite{dougherty1992}, where members need ``trading zones" \cite{galison1997}, a ``third space" \cite{sanders2012}, a notion of ``other-ness" \cite{malone2022}, or common ground \cite{pennington2011,repko2012,plaisant2018} to interact within. Communication-based challenges can be targeted more effectively and immediately, while interdisciplinary-based challenges need deeper interventions to change underlying cultures and unhelpful perceptions and mindsets. 

\begin{figure}[h]
\includegraphics[width=\linewidth]{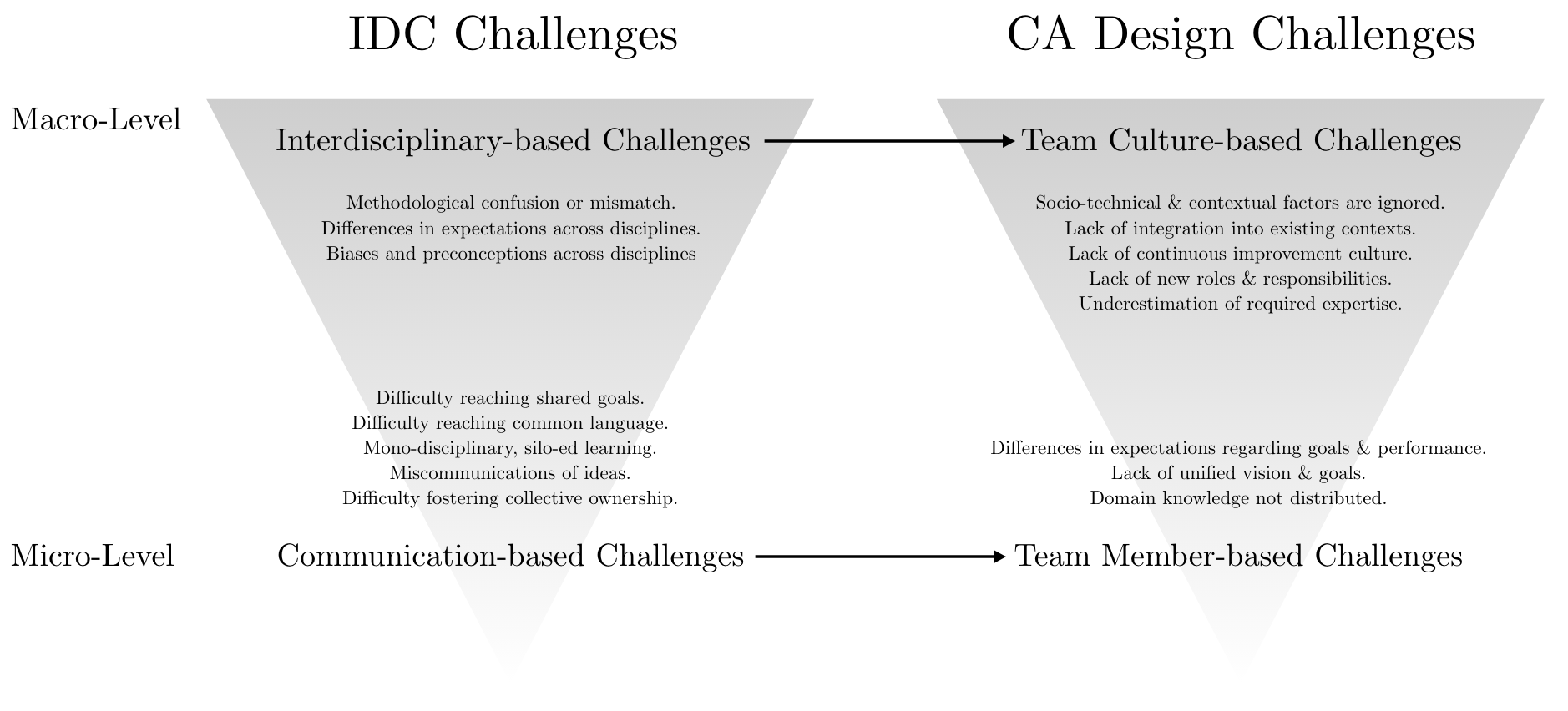}
\caption{A mapping between IDC and CA design challenges taxonomies.} \label{fig:mapping}
\end{figure}

\bigskip
\noindent Returning to the types of CA design challenges, we propose their division into \emph{CA design team culture-based challenges} and \emph{CA design team member-based challenges}, similar to the distinction of these types of challenges in IDC. While the prior need broader, longer-term strategies targeting team cultures and mindsets, the latter focus more on teamwork in general, and especially communication and individuals' perceptions of the project and its parameters and participants. Fig. \ref{fig:mapping} highlights these challenge taxonomies and their mappings.

\section{Strategies for Overcoming Conversational Agent Design Challenges}

The previous sections reviewed CA design challenges and broadly taxonomized the challenges into two groups: those arising from unhelpful team cultures and practitioners’ mindsets and attitudes, which is synonymous to different disciplines coming together, and those arising from communication issues on a team-member level, synonymous to well-documented communication challenges in interdisciplinary collaborations. As such, the proposed hypothesis is that strategies to overcome the challenges of interdisciplinary collaboration could prove effective in addressing, or at least partially addressing, some of the conversation agent design challenges mentioned.  

\bigskip
\noindent Team member challenges are heavily communication-based as discussed earlier, and can be addressed through establishing common ground by unifying goals, expectations, language and knowledge within a team. Team culture challenges can arise from the coming together of different cultures and disciplines, and can be addressed by reaching harmony through a suitable team culture or environment, which in turn can be achieved by having role clarity and appropriate practices and traditions. Each strategy encompasses a variety of actions and activities to be conducted in order to effectively adopt the corresponding strategy. These are summarised in the last segment of this section.

\subsubsection{Reaching Common Ground through Communication - Boundary Objects for Unified Language and Knowledge}

Research confirms that agreeing on common goals, research questions, theoretical frameworks, and language definitions is key to achieving synthesis and having a successful project \cite{defila2006}. Additionally, setting expectations early on \cite{stephenson2010} regarding individual stakeholders' goals and how those align with collective team goals \cite{matthews2011} is critical so that all members can begin to work together. Creating boundary objects, sometimes referred to as intermediate objects \cite{ocnarescu2011} or knowledge integrators \cite{lewandowski2021}, is a widely used technique in the field of Computer-Supported Collaborative Work (CSCW) to help reach a shared understanding or establish a shared model of reality between participants with differing perspectives or backgrounds \cite{pennington2010,cabitza2012,arrighi2019}. In interdisciplinary settings especially, the object acts as a tangible affordance of the knowledge produced, shared, and consumed throughout the activity \cite{pennington2010,cabitza2012,arrighi2019}, a form of documentation and basis for future-decision making, as well as a medium for discussions and conversations \cite{vines2013}. Similarly, design workshops also commonly revolve around the creation and/or use of such boundary objects \cite{vines2013,steen2013,jonas2018}. Boundary objects are visual in nature, but they can be digital or physical, and 2-dimensional or 3-dimensional. These objects can be closed objects that present or prescript information, such as models or design plans. Conversely, they can be open objects that act as inspiration or encourage interpretation and discussions such as sketches, decks of cards, and tangible objects \cite{redaelli2018,paavola2019}. Tangible objects and toolkits (such as clay figurines, decks of cards, wooden blocks, sticky notes, etc.) have been found to be suitable for quick changes and fast evaluation \cite{boujut2003}.  

\bigskip
\noindent Another strength of boundary objects is their ability to act as a form of documentation. It is frequently cited that a significant limitation of design processes is the lack of documentation of and reflection on the learnings and knowledge produced, and the design decisions taken throughout them \cite{sleeswijk2018,koesten2019}. This lack of documentation can lead to data going missing, and confusion over team roles, expected outcomes, and the processes taking place -- especially in trans-disciplinary settings \cite{sleeswijk2018,koesten2019}. In AI-based systems, where practitioners also complain of a lack of documentation \cite{zhang2020}, this can be especially important for transparency and accountability \cite{custis2021,fair2021}. Instead of creating dedicated documentation such as Algorithmic Impact Assessments \cite{reisman2018}, Data Privacy Impact Assessments \cite{selbst2019}, Social Impact Assessments \cite{edwards2016}, Privacy Impact Assesments \cite{oetzel2014} and Data Ethics Impact Assessments \cite{floridi2018}, which come with a range of limitations and criticisms \cite{metcalf2021,moss2021}, boundary objects created during the design process can double as documentation without posing extra work on practitioners.

\bigskip
\noindent \cite{invision2021} also reported that effective interdisciplinary organizations had a unified repository for design-based, code-based, and user-based knowledge. Deciding on a unified knowledge repository or a location for approved ideas and documents is crucial for smooth IDC \cite{Rosenberg2011,WWDC17,kelly2018}; almost like the engagement party prior to the upcoming disciplines' wedding \cite{crease2010}. Industry professionals have also called for establishing a common language between designers and developers \cite{frost2021}. In the space of designing AI-based systems (such as CAs), designers have expressed that ``[establishing] a shared understanding of the AI system" \cite{yildirim2022}~[p.5] was a critical first step that they spent a significant amount of time on in collaboration with data scientists and other stakeholders, citing the need for a shared language or boundary object. AI practitioners especially have expressed a need for boundary objects in their work, especially to aid with collaboration and interdisciplinary communication and establish a common language \cite{lee2020,yildirim2022}. \cite{yildirim2023} specifically recommend creating boundary objects that can be used by different stakeholders and team members. The use of boundary objects has also been called for \cite{lewandowski2021} and explored in the space of CA design, including whether CAs themselves can act as boundary objects \cite{castle2020}.

\subsubsection{Reaching Harmony through Team Culture - Mindsets \& Practices}

Clarifying team roles and where they overlap is critical for understanding exactly where, when, how, and between whom collaboration will take place \cite{antonic2021,steen2013}. Even in the specific context of CA design, \cite{heuer2022} recommend aiming to achieve role and vision clarity early on. Creating the right environment plays a vital role in helping team members better understand their thoughts and feelings \cite{steen2013}, helping them to immerse themselves in different disciplines \cite{stokolos2008,strober2010,adams2018} and helping them develop comfort and a psychological sense of safety. This state of comfort and safety is needed to effectively participate and reach the height of creative, innovative and problem-solving capacities \cite{stephenson2010}. These feelings of safety and comfort can also help foster cooperation, instead of feeling the need to compete, which tends to cause more conflicts and friction \cite{antonic2021}. Having open, informal group conversations that are supportive of focus shifts, off-topic branches, and low coherence can also help overcome design fixation, enhance creative ideation, and bridge gaps between seemingly unrelated ideas \cite{menning2018}. It is therefore important to encourage open mindedness and open communication to benefit from differing perspectives and elicit the requirements and constraints of different departments as early as possible \cite{winowiecki2011,Khamdamova2016}. It is also important to support sharing work-in-progress, as this can lead to team members being less apprehensive about sharing their ideas and concepts and can aid group- and self-reflection about how to improve the process used \cite{agrawala2018}. 

\bigskip
\noindent It is also crucial that each member of each team understand how important each other member of every other team is and what role they play in the bigger picture \cite{norman2002,Rosenberg2011}. By understanding and respecting the fact that each discipline has a different perspective and different requirements and constraints, and that they are all `correct', interdisciplinary teams can reach peak effectiveness and creativity. In order to create a synergy and be able to collect collective consultations and plans from the outset, instead of making costly modifications at a later stage based on the input of an isolated team that was consulted too late \cite{norman2002}, the team must think of themselves as one unit. They should thoroughly discuss and collaborate on all deliverables before moving them to stakeholders outside the team \cite{Khamdamova2016}. These practices help build a sturdy ``social fabric" which is crucial for developing relationships \cite{stephenson2010} and building trust \cite{crease2010,mcbee2017}. There must be representatives from each team or discipline present at all stages and they must learn to listen and satisfy the needs of each other in addition to the needs of users \cite{Rosenberg2011,Khamdamova2016}. 

\bigskip
\noindent \textbf{Honorable Mention: Value-Sensitivity as a Catalyst for Harmony} While slightly outside the scope of this study, it is worth mentioning that focusing on interdisciplinary team members’ values can be a unifying factor to help collaboration. Having common values or philosophies is cited as being helpful to interdisciplinary interactions \cite{nancarrow2013}. It can also help smooth over or consolidate discipline-based differences; especially as some overarching values are easy to agree on. Using techniques borrowed from the field of Value-Sensitive Design \cite{friedman2002} to help identify and find strategies to respect different stakeholders’ values while creating the technology might help provide common ground for the team to agree upon and move forward. Sometimes team members might even agree on what should happen without agreeing on the reasons why \cite{friedmanbook}~[p.48]. Additionally, ``value transparency" \cite{park2022} can prove very useful for IDC. This means making explicit the values that decision-makers prioritized or respected when making decisions as a way of knowing why decisions were made, if people cannot know how they were made. Value transparency is arguably easier to achieve than other forms of transparency and could help resolve stakeholder tensions despite the existence of value conflicts between them, as well as help establish common ground and stakeholder connections through shared values \cite{park2022,michalke2022}. These value-based approaches are being more common in the space of designing AI-based systems \cite{umbrello2021} and are slowly making their way into the space of CA design as well \cite{wambsganss2021}. 

\subsection{Future Empirical Work and Open Problems}
The work conducted by this study is conceptual in nature and while it takes an important step in the right direction, future empirical work is needed to support its claims. This future work can make several critical contributions, which are outlined below:
\begin{itemize}
    \item Examining and establishing a link of causality between CA design challenges and IDC challenges, i.e., whether this causality exists, within which challenges, to what extent, and what other sources are contributing to different CA design challenges.
    \item Determining whether a direct mapping exists between different individual CA design challenges and corresponding IDC challenges.
    \item Testing for an empirical correlation between the established CA design challenge taxonomy and a corresponding taxonomy of IDC challenges (such as communication-based versus culture or discipline-based) beyond a conceptual correlation. 
    \item Validating the proposed strategies by observing their effect on different CA design challenges.
\end{itemize}

\section{Conclusion}
The aim of this opinion paper was to suggest an exploration of the links between CA design challenges presented by \cite{heuer2022} in their scoping review and IDC challenges. While solely conceptual, the goal of this work is to inspire more empirical investigations to evaluate the strategies' effectiveness and investigate a possible link between CA design challenges and IDC, potentially paving the way for new strategies and solutions for emerging challenges in the developing space of CA design.

\begin{acks}
This project was partially funded by the Leverhulme Trust through the Leverhulme Centre for the Future of Intelligence [RC-2015-067]. This work was also supported by The Alan Turing Institute’s Enrichment Scheme.
\end{acks}
\newpage
\bibliographystyle{ACM-Reference-Format}
\bibliography{main}

\end{document}